\documentstyle[prc,preprint,aps,12pt]{revtex}
\def\half{{{\textstyle {1\over2}}}}
\def\part#1{{ \partial\over{\partial#1} }}

\begin{document}
\draft
\title {Effects of angular momentum projection on the nuclear partition\\ 
function and the observation of the giant dipole resonance\\
in hot nuclei}
\author{W.E.~Ormand$^{a,b,c}$, P.F.~Bortignon$^{a}$, R.A.~Broglia$^{a,d}$}
\address{$^a$Dipartimento di Fisica, Universit\`a di Milano, and 
Istituto Nazionale di Fisica Nucleare sezione di Milano,
Via Celoria 16, 20133 Milano, Italy\\
$^b$Physics Division, Oak Ridge National Laboratory, P.O. Box 2008, 
MS-6373 Building 6003,\\ Oak Ridge, TN 37831-6373 USA\\
Louisiana State University, Baton Rouge, LA 70803-4001 \\
$^d$The Niels Bohr Institute, University of Copenhagen,
Blegdamsvej 17, \\ DK-2100 Copenhagen \O, Denmark}
\maketitle
\begin{abstract}
\noindent
Procedures for projecting angular momentum in a model describing a hot 
nucleus that takes into account large-amplitude quadrupole fluctuations 
are discussed.
Particular attention is paid to the effect angular-momentum projection 
has on the observables associated with the $\gamma$-decay of the 
giant-dipole resonance (GDR). We also elaborate on which of the
different projection methods provides the best overall 
description of the GDR, including
angular distributions. The main consequence of angular-momentum
projection is the appearance of an effective 
volume element in the integrals associated with the thermal average of
the physical observables. This effective volume element is controlled by 
the value of the moments of inertia of the system. 
In the limit of rigid-body moments of inertia, the effective volume element is 
found to differ only slightly from the volume element associated with the 
normalization of the five-dimensional quadrupole oscillator wavefunction
in the $\beta$, $\gamma$, and $\Omega$ space, namely 
${\cal D}[\alpha]=\beta^4d\beta \sin(3\gamma)d\gamma d\Omega$. In the 
limit of irrotational flow moments of inertia, the leading behavior in the 
$\beta$ degree of freedom is ${\cal D}[\alpha]\propto \beta d\beta$. 
\end{abstract}

\section{Introduction}
\label{sec:intro}
It is well known that when a nucleus is observed at finite excitation energy, 
large-amplitude thermal fluctuations can play an important role. 
An example is provided by the cross section of the giant-dipole resonance 
(GDR) in a hot nucleus~\cite{r:Gal85,r:Pac88,r:Alh88}, which 
may be written as a weighted average over the deformation parameters 
$\alpha_{\lambda\mu}$,
\begin{equation}
\sigma(E) = Z^{-1} \int {\cal D}[\alpha] \sigma(\alpha;E) 
{\rm e}^{-F(T,\alpha )/T}.
\label{e:eq1}
\end{equation}
Here $E$ is the photon energy, $\sigma(\alpha;E)$ is the 
deformation-dependent  cross section of the GDR evaluated in the
laboratory frame. The quantity $F$ 
represents the free energy of the system at temperature $T$, and
\begin{equation}
Z=\int {\cal D}[\alpha] \exp(-F/T),
\end{equation}
is the partition function.
Eq.~(1) is equivalent to the single-time slice limit, or static-path 
approximation (SPA), of the path-integral formulation of 
Refs.~\cite{r:Lau88,r:Joh91}, and should be understood as an adiabatic limit: 
i.e., under the assumption that the time scale for 
sampling the quadrupole shape deformations 
is long compared to the change in the dipole frequency associated with the
shape fluctuations. For a description of the effects of nonadiabatic thermal 
fluctuations on the GDR, see Refs.~\cite{r:Alh89,r:Orm90}. 

In the case of the GDR, the strongest coupling is to the quadrupole 
degrees of freedom, which are defined by the quadrupole deformation 
parameters $\beta$ and $\gamma$, and the Euler angles 
$(\phi,\theta,\psi)$\footnote{The Euler angles and rotation matricies are 
defined using the convention of Edmonds~\cite{r:Edm60}. At an arbitrary
orientation, the $\alpha_{2\mu}$ are given by $\alpha_{2\mu}=\sum_m a_m
D^2_{\mu m}$, with $a_0=\beta\cos\gamma$, 
$a_2=a_{-2}=\beta\sin\gamma/\sqrt{2}$, and $a_1=a_{-1}=0$.}
defining the intrinsic orientation of the system. 
The volume element ${\cal D}[\alpha]$ has been the source of  
controversy in the past~\cite{r:Gal85,r:Pac88,r:Alh88}. 
With the choices being between
${\cal D}^{(1)}[\alpha]=\beta^4d\beta \sin(3\gamma)d\gamma d\Omega$ 
\cite{r:Lau88}, with 
$d\Omega=\sin \theta d\theta d\phi d\psi$ and 
${\cal D}^{(2)}[\alpha]=\beta d\beta d\gamma d\Omega$ 
\cite{r:Gal85,r:Pac88}\footnote{In Refs.~\cite{r:Gal85,r:Pac88}, 
the integration over $d\Omega$ was not performed}.
The first choice arises naturally 
in the quantization of the five-dimensional quadrupole oscillator, or Bohr 
Hamiltonian~\cite{r:Boh75}, and the static-path approximation to the nuclear
partition function~\cite{r:Lau88} in the limit of that the residual interaction 
contains only quadrupole-quadrupole terms. 
The second is a generalization to
non-axially symmetric quadrupole deformations of the results obtained by
Bertsch~\cite{r:Ber80}, where the number of states per unit interval of
$\beta$, was found to be proportional to $d\beta$. 

In this work, three methods for projecting angular momentum 
onto a nuclear system at finite temperature are investigated. 
In the standard approach, which was used in most previous theoretical studies 
of the GDR in hot nuclei, angular momentum is projected on average by 
using a constant rotational frequency, and the GDR is basically
described via Eq.~(\ref{e:eq1}) with ${\cal D}[\alpha ]={\cal D}^{(1)}[\alpha ]$.
An important drawback of this method occurs at high spin, where the free
energy exhibits a fission saddle-point beyond which $F\rightarrow -\infty$ 
and Eq.~(\ref{e:eq1}) is unbound. In order to overcome this
difficulty in connection with the study of the Jacobi-shape phase
transition, which is likely to take place for values of angular momentum
close to those leading to fission, total angular momentum projection, 
as is outlined in the next section, was introduced in Ref.~\cite{r:Alh93}.

In the present work, we examine the
consequences of angular momentum projection on the 
GDR, and conclude that it
leads to an effective volume element that is determined by the moments of
inertia. In the limit of rigid-body moments of inertia, the effective
volume element is found to differ only slightly from ${\cal
D}^{(1)}[\alpha]$, while in the limit of irrotational flow moments of
inertia, the leading behavior in $\beta$ for the effective volume element 
is $\beta d\beta$. The third approach discussed here is to project only
the $z$-component of the angular momentum in the calculation of 
the partition function. It is 
found that this method leads to strength functions and
full-width-at-half-maxima that are very similar to the second projection
method at low spins, but it gives a better overall description of non-scalar 
observables, such as the angular distribution $a_2$-coefficient.

The general outline of this paper is as follows. 
In Section~\ref{sec:method}, the
three methods of projecting angular momentum are derived and discussed.
In Section~\ref{sec:observ}, 
the application of projection methods to the GDR is 
discussed, while comparisons between the methods are shown as a function
of angular momentum for the nuclei $^{106}$Sn and $^{208}$Pb in 
Section~\ref{sec:compare}. Concluding remarks are given in 
Section~\ref{sec:conc}.

\section{Angular Momentum Projection Methods}
\label{sec:method}
The intrinsic cross section $\sigma (\alpha ;E)$ of the GDR and 
the free energy $F$ are naturally evaluated in the cranking model at a 
fixed frequency $\omega$, while the (external) cross section should be 
evaluated at a fixed angular momentum $J$.
The transformation from frequency to angular momentum may be performed by 
adjusting the frequency to give the correct angular momentum value.
If this is done independently for each quadrupole deformation and 
orientation, one is lead to use an effective volume element in 
Eq.~(\ref{e:eq1}). To see this, we begin with the partition function
at fixed frequency and temperature,
\begin{equation}
Z(\omega ,T)= {\rm Tr} \left( {\rm e}^{-(H-\omega J_z)/T}\right) = 
\int {\cal D}[\alpha] {\rm e}^{-F(T,\alpha ,\omega )/T}.
\end{equation}
Except for a prefactor to the integral, we note that this expression is 
obtained in the static-path approximation to a quadrupole deformed 
nucleus at finite temperature~\cite{r:Lau88}. 
The volume element is given by 
${\cal D}[\alpha]={\cal D}[\alpha]^{(1)}$ and 
${\rm e}^{-F(T,\alpha ,\omega )/T}$ is the probability distribution at 
fixed frequency.
The partition function for a fixed value $M$ of the $z$-component of the 
angular momentum is given by the inverse Laplace transform with respect 
to the frequency
\begin{equation}
Z_M(T)= {\rm Tr}_M \left( {\rm e}^{-H/T}\right) = 
\frac{1}{2\pi iT} \int_{-i\infty}^{i\infty} d\omega
{\rm e}^{-\omega M/T} Z(\omega ,T).
\end{equation}
The partition function at fixed angular momentum $J$ is then obtained by 
differentiation with respect to $M$ via
\begin{eqnarray}
Z_J(T) &=& {\rm Tr}_J \left( {\rm e}^{-H/T}\right)
= (2J+1)[ Z_{M=J}(E) - Z_{M=J+1}(E) ] \nonumber\\
&\approx & -(2J+1) \frac{\partial}{\partial M} Z_M(E)
\bigg\vert_{M=J+\half} \nonumber\\
&=&\frac{2J+1}{2\pi iT^2} \int_{-i\infty}^{i\infty} d\omega \omega
{\rm e}^{-\omega (J+\half)/T}Z(\omega ,T)\nonumber \\
&=& \frac{2J+1}{2\pi iT^2} 
\int {\cal D}[\alpha] \left [ \int_{-i\infty}^{i\infty} d\omega\omega
{\rm e}^{  -\left(\omega(J+\half)+F(T,\alpha,\omega)\right)/T}\right ],
\label{e:eq4}
\end{eqnarray}

In the last expression of Eq.(\ref{e:eq4}), 
the quantity inside the square bracket may be 
identified with the effective 
probability distribution for the deformation characterized by the 
parameters $\alpha$ at fixed spin.
The integration over $\omega$ is usually approximated by applying the 
saddle-point technique~\cite{r:Arfkin}. 
In this regard, there are two choices: the application of the saddle-point 
approximation before or after the integration over deformation degrees of 
freedom. In the latter case, the saddle-point condition is
\begin{equation}
J+\half = T\part{\omega} \ln Z(\omega,T) \bigg\vert_{\omega=\omega_J}, 
\label{e:eqrot}
\end{equation}
leading to a constant rotational frequency, $\omega_J$, applied at all 
deformations, and the fixed angular momentum partition function is given by
\begin{equation}
Z_\omega(J,T) = \frac{(2J+1)}{T^2(2\pi{\rm D})^{1/2}}\omega_J
{\rm e}^{-\omega_J(J+1/2)/T} 
\int {\cal D}[\alpha] {\rm e}^{-F(T,\alpha,\omega_J)},
\end{equation}
where ${\rm D}=\frac{\partial^2}{\partial\omega^2}
\ln Z(\omega,T)\vert_{\omega=\omega_J}$.
This approach applies a 
constraint so that the expectation value of the angular momentum is, 
on average, equal to $J$. As such, at higher spins, large variations in
the projected angular momentum, which can be measured by computing the
variance of $J$, can be expected.

In the case of applying the saddle-point integration before 
integrating over the deformation degrees of freedom, one obtains a rotational 
frequency, $\omega_{\alpha,J}$, for each deformation and orientation that
is given by
\begin{equation}
J+\half = -\part{\omega} F(T,\alpha,\omega) 
\bigg\vert_{\omega=\omega_{\alpha,J}}.
\label{e:eq5}
\end{equation}
It is an open question as to which choice of performing the integration 
over $\omega$ leads to a better evaluation 
of the nuclear partition function, and, by extension, the expectation value of 
spin-dependent observables. However, we note that since the nuclear free energy 
is a rotational invariant, it must be 
at least a quadratic function of the rotational frequency. Consequently, to
lowest order, the free energy may be written as
\begin{eqnarray}
\label{e:eq6}
F(T,\alpha,\omega) &=& F(T,\alpha,\omega=0)-\frac{1}{2}\omega^2(I_1\cos^2\psi\sin^2\theta + 
I_2\sin^2\psi\sin^2\theta +I_3\cos^2\theta),\\
&=&F(T,\alpha,\omega=0)-\frac{1}{2}\omega^2
{\cal I}(\beta,\gamma,\theta,\psi),
\end{eqnarray}
where the $I_k$ are the nuclear moments of inertia.
At this point, we note that as long as there are no quartic terms in the free 
energy, Eq.(\ref{e:eq4}) can be evaluated analytically,
and the saddle-point integral applied before the integration over the 
quadrupole degrees of freedom, as indicated in Eq.(\ref{e:eq5}), is exact.
Therefore, one may write the partition function as
\begin{equation}
Z(J,T) = \frac{(2J+1)^2}{2 (2\pi T^3)^{1/2}} \int 
\frac{{\cal D}[\alpha]}{{\cal I}(\beta,\gamma,\theta,\psi)^{3/2}} 
\exp\left [ -F(T,\alpha,J)\big/T\right ].
\label{e:eq7}
\end{equation}
where
$F(T,\alpha,J) = F(T,\alpha,\omega =0) + 
(J+\half)^2/2{\cal I}(\beta,\gamma,\theta,\psi)$. Note that the 
saddle-point frequency of Eq.~(\ref{e:eq5}) corresponds to
\begin{equation}
\omega_{\alpha,J}=\frac{J+\half}{{\cal I}(\beta,\gamma,\theta,\psi)}. 
\label{e:saddle_j}
\end{equation}

From Eq.(\ref{e:eq7}), it is apparent that the moments of inertia 
define an effective volume element, 
${\cal D}^{eff}[\alpha] ={\cal D}[\alpha]/{\cal I}(\beta,\gamma,\theta,\psi)^{3/2}$.
For nuclei, there are two limiting cases for the moments of
inertia that are of interest. 
The first being when the angular momentum is 
built up entirely from the collective degrees of freedom, at which point the 
moments of inertia would correspond to the irrotational-flow 
values: 
$I_k^{irrot} = I_0\beta^2\sin^2(\gamma + 
2\pi k/3)$, where 
$I_0= (3/2\pi)MR_0^2$, and the effective volume element 
is proportional to $\beta d\beta$ and is given by
\begin{equation}
\frac{\beta d\beta\sin(3\gamma)d\gamma d\Omega}
{[I_0(\sin^2(\gamma+2\pi/3)\cos^2\psi\sin^2\theta+
\sin^2(\gamma+4\pi/3)\sin^2\psi\sin^2\theta+
\sin^2(\gamma)\cos^2\theta)]^{3/2}}, 
\end{equation} 
which differs sharply from ${\cal D}^{(1)}[\alpha]$. Indeed, 
since there are shapes for which $I_k^{irrot}=0$, there are orientations
in which ${\cal D}^{eff}$ diverges to infinity. These apparent
divergences, however, are brought under control by a similar factor in the
exponential of the free energy\\ 
$\exp[-(J+\half)^2/2{\cal I}(\beta,\gamma,\theta,\psi)]$. 

The second limit of interest is the case where the nuclear moments of 
inertia assume the rigid-body values,  
$I_k^{rigid} \approx I_0 
\left[1-\sqrt{5/4\pi}\beta\cos(\gamma +2\pi k/3)\right]$
with $I_0=(2/5)MR_0^2$. In this case, the effective volume element is
similar to ${\cal D}^{(1)}[\alpha]$.
Shown in Fig.~\ref{f:fig1} is the contour plot of 
$I_0^{3/2} {\cal D}^{eff}[\alpha]/{\cal D}^{(1)}[\alpha] = 
\left[I_0/{\cal I}(\beta,\gamma,\theta,\psi)\right]^{3/2}$ for 
rigid-body moments of inertia with Euler angles
$\phi=0$ and $\theta=\psi=\pi/4$.
As can be seen in the figure, 
the effective volume element obtained with the 
rigid-body moments of inertia differs from ${\cal D}^{(1)}[\alpha]$ 
for $\beta\le 1.0$ by at most a
factor of 2.5. This difference is not readily observable 
in the partition function (or observables of the GDR) for 
temperatures less than 2~MeV because the integral in 
Eq.(\ref{e:eq7}) is dominated by the free energy in the exponential. 
In fact, for such large values of $\beta$ the free energy is often larger  
than the minimum value by 20~MeV or more, and, 
consequently, is suppressed by the exponential factor.

As a further illustration of the effects of angular momentum
projection, density distributions for the effective weight function 
$W={\cal D}^{eff}\exp[-F/T]$ are plotted in Fig.~\ref{f:fig2} for the nucleus
$^{106}$Sn at temperatures $T=1$ and 3~MeV and zero angular  
momentum for irrotational flow and rigid-body moments of inertia as well
for the fixed rotational frequency ($\omega$) method. The free energy
was evaluated using the temperature-dependent liquid-drop
parameterization of Guet {\it et \l.}, which is based on finite temperature,
extended Thomas-Fermi calculations~\cite{r:Gue88}. In the figure, the darker
shaded areas reflect the largest weight. Note that the weight functions
obtained for the angular momentum projected case using rigid-body moments of
inertia are very similar in shape to the fixed $\omega$ method. There is, 
however, an important difference. In the fixed $\omega$ method,
each of the three regions have the same relative weight, whereas, in the
case of projected angular momentum, the weight function is asymmetric in the
three regions. This is due to an angular anisotropy in the factor 
${\cal I}^{-3/2}(\beta,\gamma,\theta,\psi)$, which arises, in part, from
fluctuations of the rotational frequency while projecting angular
momentum. Although this factor essentially allows for an exact 
computation of the partition function at the mean field level via 
Eq.~(\ref{e:eq7}), the anisotropy in the effective volume element 
can have important consequences for observables of the GDR. In particular, 
it will be shown in the Section~\ref{sec:compare} 
that although the total GDR strength
function is essentially unaffected, 
the angular distribution $a_2$-coefficient
(defined in the next section) is, and is one motivation for
using the projection method outlined below for just projecting $J_z$.

The nuclear moments of inertia can be evaluated from
cranked Nilsson-Strutinsky calculations, which, at high temperature, tend
towards the rigid-body limit. At low temperature, however, significant
corrections due to shell structure and the pairing interaction may be
present, and the moments of inertia may correspond to a mixture of the two
limits. In practice, however, the moments of inertia never correspond
solely to the irrotational limit, as this would preclude angular momentum 
built up of non-collective single-particle excitations. 

An important difference between these two methods of projecting angular momentum
has to do with the presence of a fission barrier at finite spin. In the fixed 
rotational frequency scheme, there is always a saddle point in the free energy 
along the prolate collective axes for $\omega_{rot} > 0$. The position and height of 
the saddle point naturally decreases with increasing rotational frequency, 
as is illustrated in Fig.~\ref{f:fig3}, where the
free energy in the fixed rotational frequency approach is plotted for
the nucleus $^{106}$Sn at a temperature $T=2.0$~MeV for 
rotational frequencies $\omega_{rot}= 0.0$, 0.56, 1.02, and 1.31~MeV, which 
approximately correspond to angular momentum values 0, 20, 40, and $60\hbar$,
respectively. 

Given the presence of the fission barrier, care must be taken when 
computing the partition function and accounting for the effects of  
large-amplitude thermal fluctuations on the GDR since beyond the 
saddle point the free energy decreases to $-\infty$, and from a technical 
point of view, the integral over all quadrupole degrees of freedom is 
unbound. In this case, 
the integration over $\beta$ must be limited to values below the saddle 
point. On the other hand, using the method of Eq.~(\ref{e:eq7}),
there is no saddle point in the 
free energy, and the integration over the quadrupole degrees of freedom 
is always bound. In addition, as a function of angular momentum, the 
mean-field nuclear shape displays three distinct characteristics instead of two 
as in the case of fixed rotational frequency. At low spin and temperature,
the shape is governed by the ground state, equilibrium shape. For example, 
for a deformed nucleus, a prolate shape undergoing a collective rotation about an axis perpendicular 
to the symmetry axis. As the spin and temperature increase, the nucleus  
makes a transition to an oblate shape rotating about the symmetry axis, while
at very high spins, the nucleus undergoes the Jacobi transition, which is
characterized by  
a sudden change to a prolate shape with a very large deformation rotating 
about an axis perpendicular to the two symmetry axes. Note that 
only the first two regimes are present in the fixed rotational frequency method 
of projecting the angular momentum. 

We conclude this section by noting that instead of projecting the total 
angular momentum onto the partition function via Eq.~(\ref{e:eq4}), it may be 
more appropriate in some cases to approximate total $J$ projection with
just projecting the $z$-component of the angular momentum, $M$. In this case,
the partition function becomes, 
\begin{equation}
 Z_M =  
\frac{1}{\sqrt{2\pi T}}\int \frac{{\cal D}[\alpha]}{{\cal I}(\beta,\gamma,\theta,\psi)^{1/2}} 
\exp\left[ -F(T,\alpha,M)/T \right] ,
\label{e:z_proj}
\end{equation} 
where the free energy is given by
\begin{equation}
F(T,\alpha,M) = F(T,\alpha,\omega=0) + 
M^2/2{\cal I}(\beta,\gamma,\theta,\psi),
\end{equation}
and the saddle-point frequency is given by 
$\omega_{\alpha,M}=M/{\cal I}(\beta,\gamma,\theta,\psi)$.

\section{GDR OBSERVABLES}
\label{sec:observ}
The expectation value of a scalar observable, such as the total cross section 
of the giant-dipole resonance follows in a similar manner as the partition 
function. This is primarily because scalar observables commute with the 
angular momentum projection operator. In particular, the response function of 
the giant-dipole is proportional to the Fourier transform of the ``real''-time
autocorrelation function 
\begin{equation}
\sum_i \langle i \mid (\sum_\mu d^\dagger_\mu(t) d_\mu(0)) 
{\rm e}^{-\beta\hat H} \hat P_J \mid i \rangle,
\end{equation}
where $d_\mu$ is the $\mu^{th}$-component of the isovector dipole transition 
operator and $\hat P_J$ is an angular momentum projection operator. Because of 
its high collectivity, the cross section of the GDR is often approximated by 
harmonic vibrations along the three 
principal nuclear axes with frequencies proportional to the inverse of the 
radius of each axis~\cite{r:Ner82}. Within this framework, the cross 
section for the GDR is evaluated at each deformation, orientation, and 
rotational frequency, and is denoted by $\sigma(\alpha,\omega;E)$, and is 
given by Eq.~(35) of Ref.~\cite{r:Orm97}. The total cross section,
including thermal averaging, is then 
\begin{equation}
\sigma_J(E)= Z^{-1}(J,T) \frac{2J+1}{2\pi iT^2}
\int {\cal D}[\alpha] \left[\int_{-i\infty}^{i\infty} d\omega\omega
\sigma(\alpha,\omega;E) 
\exp\left(-(\omega(J+\half)+F(T,\alpha,\omega))\big/T\right) \right].
\label{e:gdr}
\end{equation}

Approximating Eq.~(\ref{e:gdr}) by evaluating the dipole cross section at the 
saddle-point frequency for each deformation as defined in Eq.~(\ref{e:eq5}) 
leads to 
\begin{equation}
\sigma_J(E) \approx Z^{-1}(J,T) \frac{(2J+1)^2}{2(2\pi T^3)^{1/2}}
\int {\cal D}^{eff}[\alpha]\sigma(\alpha,\omega_{\alpha,J};E)
\exp\left[ -F(T,\alpha,J)/T\right] .
\label{e:sig_j} 
\end{equation}
In the fixed rotational frequency approach, 
where the rotational frequency is defined by 
Eq.~(\ref{e:eqrot}), 
the dipole cross section is evaluated at the saddle-point 
frequency $\omega_J$ for each deformation and orientation, giving
\begin{equation}
\sigma_{\omega,J}(E) \approx Z^{-1}(\omega_J,T) \int {\cal D}^{(1)}[\alpha] 
\sigma(\alpha,\omega_J;E)\exp\left[ -F(T,\alpha,\omega_J)/T\right] .
\label{e:sig_om}
\end{equation}
Finally, when only the $z$-component of the angular momentum is
projected, the GDR cross section is
\begin{equation}
\sigma_M(E) \approx Z^{-1}_M, \frac{1}{\sqrt{2\pi T}}
\int \frac{{\cal D}^{(1)}[\alpha]}{{\cal I}(\beta,\gamma,\theta,\psi)^{1/2}}
\sigma(\alpha,\omega_{\alpha,M};E)
\exp\left[ -F(T,\alpha,M)/T\right] .
\label{e:sig_m}
\end{equation}

In regards to non-scalar observables, we note that much of the formalism 
described here is not adequate because of the fact that these observables 
do not 
commute with operators used to project the angular momentum, and it 
may be necessary to make further approximations. 

An important non-scalar observable associated with the 
the giant-dipole resonance is the angular distribution  
$a_2$-coefficient defined as
\begin{equation}
\sigma(E,\theta) = \sigma(E)\left[1+a_2(E) P_2(cos\theta)\right],
\end{equation}
where $\theta$ is the angle between the observed GDR $\gamma$-ray and the 
polarized spin direction. In heavy-ion fusion experiments with high angular 
momenta, the $z$-component of the angular momentum is large ($\approx J$), 
and lies in a plane perpendicular to the incident beam direction. 
The angular distribution is often measured with the angle $\theta$ being 
between the observed gamma-ray and the incident beam direction. In the 
large $J$ limit, the $a_2(E)$-coefficient is given by
\begin{equation}
a_2(E) = \frac{1}{\sigma_{tot}(E)}
\left[ \sigma_0(E) - \frac{1}{2}(\sigma_1(E) + \sigma_{-1}(E))\right],
\end{equation}
where $\sigma_\mu(E)$ is the cross section for the 
$\mu^{th}$-component of the 
GDR and $\sigma_{tot}=\sum_\mu \sigma_\mu$ is the total cross section. 
In the 
standard approach, each of the $\mu$-components of the GDR strength
function are computed using Eqs.~(\ref{e:sig_j})-(\ref{e:sig_m}) for
each of the three projection methods, respectively.

\section{COMPARISON OF THE METHODS}
\label{sec:compare}
In this section, we compare the results obtained with the three different 
projections methods as a function of temperature and angular momentum 
for the nuclei $^{106}$Sn and $^{208}$Pb. 
The free energies were calculated using the 
Nilsson-Strutinsky procedure \cite{r:Str66,r:Bra81}, using the  
Nilsson-model and liquid-drop parameters given in Refs.~\cite{r:Nil69} and 
\cite{r:Gue88}, respectively. In addition, the moments of inertia were
taken to be the rigid-body values with shell corrections applied
as described in Ref.~\cite{r:Orm97}. For
the most part, the two nuclei are different in character, as the shell
corrections in $^{106}$Sn are essentially zero. This is in sharp contrast 
with $^{208}$Pb, where strong shell corrections are present at low
temperature that favor the spherical shape.

For the cross section of the giant dipole resonance, 
we use an extension of the quantal oscillator
of Ref.~\cite{r:Ner82} to include arbitrary orientations and 
quadrupole shapes, which is described in Ref.~\cite{r:Orm97}. 
For a given quadrupole shape, the principal GDR modes were taken to have
energy $E_k=E_0\exp[-\sqrt{5/4\pi}\beta\cos(\gamma + 2\pi k/3)]$.
Following experimental findings, the width of each mode 
associated with a given deformation was assumed to depend on the dipole 
energy by $\Gamma_k = \Gamma_0 (E_k/E_0)^\delta$~\cite{r:Car74}, with 
and $\delta=1.8$. As in Ref.~\cite{r:Orm97}, the parameters 
$E_0$ and $\Gamma_0$ are taken from ground
state data and are $E_0=14.99$~MeV and $\Gamma_0=5.0$~MeV for
$^{106}$Sn and $E_0=13.65$~MeV and $\Gamma_0=4.0$~MeV for $^{208}$Pb.

Shown in Fig.~\ref{f:fig4} are the cross sections and $a_2$-coefficients 
for the GDR 
in $^{106}$Sn at $T=1.6$~MeV and for $J=0$, 20, 40, and $60\hbar$ for
the three projection methods. For the most part, the total cross section 
is the same for each of the three methods with the exception that
at $J=60\hbar$ the fixed $\omega$-procedure yields a somewhat broader 
strength function. This is primarily due to the presence of the fission
barrier, which tends to limit the range of the integration and provides a
much greater weight to larger deformations along the prolate collective
axis. In regards to the $a_2$ coefficients, note that while the fixed
$\omega$ method yields a zero value for $J=0\hbar$, both the
$J$- and $J_z$-projection methods do not, although the 
$J_z$-projection value is smaller. This rather 
unphysical situation is caused by the asymmetry in the weight
function described in Section~\ref{sec:method} 
and the fact that effects due to 
fluctuations in the rotational frequency on the GDR response function
are not accounted for. On the other hand, for $J\ge 20$ all three
methods give approximately the same results. This is primarily due to
the fact that at higher spin, the asymmetry in the weight function is
governed by the exponential factor. 

By inspecting Fig.~\ref{f:fig4}, one sees that the principal effects
angular momentum has on the GDR are to broaden the total cross section
and to enhance the $a_2$-coefficient. These effects are brought
about for two reasons. First, as the angular momentum increases, the 
minimum in the free energy shifts towards the oblate non-collective axis, 
and the mean deformation increases. The three $\mu$-components then split, 
and as a result, the total cross section, which is the sum of all three 
components, broadens. For the $a_2$-coefficient, not only is there a
splitting caused by the mean deformation, but angular momentum causes
an angular anisotropy in the weight factor, primarily through the
exponential of the free energy.

Shown in Fig.~\ref{f:fig5} are the cross section and $a_2$-coefficient 
for $^{208}$Pb at $T=1.5$~MeV for $J=0$, 20, 40, and $60\hbar$ for
the three projection methods. 
As in $^{106}$Sn, the three different methods give
roughly the same total cross section, while the $J$-projection 
method yields a nonzero $a_2$-coefficient for $J=0\hbar$. 
Again all three methods yield approximately the same $a_2$-coefficient
for $J\ge 20\hbar$. From a
comparison of Figs.~\ref{f:fig4} and \ref{f:fig5}, it is apparent 
that for $^{208}$Pb both the width of the total cross section 
and the $a_2$-coefficient exhibit a weaker dependence on angular momentum 
than for $^{106}$Sn. This is primarily due to the larger moment of inertia
in Pb that leads to both a smaller
splitting between the $\mu$-components of the GDR and a smaller angular
anisotropy.

Another important difference between $^{208}$Pb and $^{120}$Sn is the 
influence of shell corrections. The most important of these are the
shell corrections to the free energy at zero angular momentum, 
which in $^{208}$Pb are quite
strong and tend to enhance the spherical shape at lower temperatures. 
In addition, because of the closed-shell structure of $^{208}$Pb, 
shell corrections also have a strong effect on the moments of inertia. 
Indeed, in Ref.~\cite{r:Orm97} it was found that shell corrections reduce 
the moment of inertia for the spherical shape by over 40\% at zero
temperature. Because of the ${\cal I}^{-3/2}$ dependence in the ``effective'' volume
element in Eq.~(\ref{e:eq7}), it might be expected that the 
shell corrections to the moment of inertia would
significantly affect the GDR strength function, as they  
appear to give a stronger preference to the spherical shape. 
We find, however, that
because of the $\beta^4$ factor in the volume element, these shell
corrections have very little effect on the 
GDR cross section beyond that produced by 
the free energy. This is exhibited in Fig.~\ref{f:fig6}, where 
the weight factor (which omits the $\sin3\gamma$ factor)
$W=\beta^4/{\cal I}^{3/2}{\rm e}^{-F/T}$ is plotted for 
oblate and prolate shapes at $T=1.0$~MeV for various combinations of
the shell corrections. In the top panel of the figure, the weight factor
is plotted including shell corrections to the free energy as well as
with and without shell corrections to the moments of inertia.
The corresponding figure without shell corrections is
shown in the bottom part of the figure. In both cases, it is seen that
the overall behavior of the weight function is governed by the exponential 
of the free energy, which is plotted in the upper right-hand panel. In 
addition, the ratio ${\cal I}_{LD}/{\cal I}_{SHL}$ is shown in the lower
left-hand panel, where it is seen that without the $\beta^4$ factor the
spherical shape would have approximately 40\% more weight when shell 
corrections to the moments of inertia are included.

At higher spins, the shell corrections to the moment of inertia 
in $^{208}$Pb only have 
a moderate effect on the total
GDR cross section and the $a_2$-coefficient. This is despite the fact that 
these shell corrections significantly reduce the moment of inertia from 
the rigid-body value for nearly spherical shapes. The principal effect
of the shell corrections would be to give to a slightly larger deformation along
oblate non-collective axis in the minimum of the total free energy. 
Note from panel (d) in Fig.~\ref{f:fig6}, however, that these  
shell corrections tend to damp out quickly for 
increasing $\beta$. On the other hand, the shell corrections to the free
energy, $F(\alpha,\omega=0)$, are quite strong, and favor the spherical 
shape, and, for the most
part are the dominating factor determining the influence of thermal 
fluctuations of the shape on the GDR. In fact, we find that at $T=1.5$~MeV,
the $a_2$-coefficient obtained at $J=60\hbar$ 
with and without shell corrections to the moment of inertia are essentially
identical.

In regards to using purely irrotational flow moments of inertia for the
calculation of the GDR properties, we note from Fig.~\ref{f:fig2} 
that not only due very small 
deformations have considerable weight, but they 
also have very small values for 
the moments of inertia. As a consequence, the saddle-point rotational
frequency can be quite large ($> 10$~MeV), which leads to unphysical
results within the model for the GDR. 

To conclude this section, we note that to some degree all three methods
lead to similar results with various degrees of problems in different
regions. Given that the $J_z$-projection
method can account for the Jacobi transition at high spin, whereas the
fixed rotational frequency procedure cannot, and that it leads to more 
reasonable $a_2$-coefficients for lowest values of angular momentum than
does the full $J$-projection method, it is likely that 
the best all around method
for describing the GDR in a hot nucleus at high spin is the $J_z$-
projection method.

\section{Conclusions}
\label{sec:conc}
We found that the choice of the method for 
projecting angular momentum 
onto the nuclear partition function can affect the sampling of 
large-amplitude thermal fluctuations of the nuclear shape. In particular, 
angular momentum projection can lead to an effective volume element that 
depends on the moments of inertia of the system. We have shown that in the 
case of irrotational-flow moments of inertia, the effective volume element 
scales as $\beta d\beta$, whereas in the rigid-body limit, 
${\cal D}^{eff}[\alpha]$ 
differs only slightly from the standard value of 
$\beta^4 d\beta cos(3\gamma ) d\gamma d\Omega$. 

In practical terms, the
three methods give essentially the same cross section for low to medium
values of angular momentum. At higher angular momentum, the fixed
rotational frequency procedure has a decreasing fission barrier that
tends to favor prolate collective shapes, but is essentially unbound, as
beyond the barrier the integrand tends to infinity. In contrast, both
the $J$- and $J_z$-projection methods have weight functions that are
always bound and exhibit the Jacobi transition at very high spin. The
primary drawback of the $J$-projection technique is that at it fails to
yield a zero valued $a_2$-coefficient at $J=0\hbar$, whereas in
the $J_z$-projection method the $a_2$-coefficient is much smaller. In
conclusion, a reasonable approach to projecting angular momentum for 
predicting features of the giant-dipole resonance in hot nuclei is 
the $J_z$-projection method.

\begin{center}
{\bf Acknowledgments}
\end{center}
Oak Ridge National Laboratory is managed for the U.S. Department of Energy
by Lockheed Martin Energy Research Corp. under contract No.
DE--AC05--96OR22464.  This work was supported in part by NSF Cooperative
agreement No. EPS~9550481, NSF Grant No. 9603006, and DOE contract
DE--FG02--96ER40985.
\newpage

\bibliographystyle{try}

\begin{figure}
\caption{Contour plot of the effective volume element ${\cal D}^{eff}[\alpha]$
normalized by the factor ${\cal D}^{(1)}[\alpha]/I_0^{3/2}$ as a function the
deformation parameters $\beta$ and $\gamma$ for the Euler angles $\phi=0$, and
$\theta=\psi=\pi/4$. The radius $\beta=1$ is indicated by the dashed line, and
the special values of $\gamma$ denoting oblate-noncollective ($\gamma=\pi/3$),
prolate collective ($\gamma=0$), oblate collective ($\gamma=-\pi/3$), and
prolate noncollective ($\gamma=-2\pi/3$) rotations are labeled.}
\label{f:fig1}
\end{figure}

\begin{figure}
\caption{Density distributions of the effective weight factor 
$W={\cal D}^{eff}[\alpha]\exp[-F/T]$ as a function of the deformation parameters
$\beta$ and $\gamma$ for the nucleus $^{106}$Sn at temperatures $T=1.0$ and
3.0~MeV and zero angular momentum for irrotational-flow (top panels)
and rigid-body (middle panels) moments of inertia. In the bottom panels, the
density distributions obtained with the fixed rotational frequency projection 
method assuming rigid-body moments of inertia are plotted.}
\label{f:fig2}
\end{figure}

\begin{figure}
\caption{The free energy for $^{106}$Sn is plotted (lower panel)
along the oblate noncollective ($\beta < 0$)
and prolate collective ($\beta > 0$) axes at a temperature of 2~MeV and 
a rotational frequency of 1.25~MeV $(\langle J\rangle\approx 55\hbar)$. In
the upper panel, the Boltzman weight factor 
$\exp[-(F-F_{eq})/T]$, where $F_{eq}$ is the minimum of the free energy below
the saddle point, is plotted.}
\label{f:fig3}
\end{figure}

\begin{figure}
\caption{Comparison between the different methods of angular momentum
projection for the cross section and $a_2$-coefficients for the GDR in
$^{106}$Sn at $T=1.6$~MeV and $J=0$ (solid line), 20 (dashed
line), 40 (dotted line), and 60 (dot-dashed line).}
\label{f:fig4}
\end{figure}

\begin{figure}
\caption{Comparison between the different methods of angular momentum
projection for the cross section and $a_2$-coefficients for the GDR in
$^{208}$Pb at $T=1.6$~MeV and $J=0$ (solid line), 20 (dashed
line), 40 (dotted line), and 60 (dot-dashed line).}
\label{f:fig5}
\end{figure}

\begin{figure}
\caption{The weight function 
$W(\beta)=\beta^4/{\cal I}^{3/2}{\rm e}^{-F/T}$ at $T=1.0$~MeV
for prolate ($\beta >0$) and oblate ($\beta < 0$) shapes. In panel (a), 
$W(\beta)$ includes shell corrections to the free energy,
$F_{SHL}$, as well as with (dotted line) and without (solid line) 
shell corrections to the moments of inertia. In panel~(b), 
the same quantities are plotted without shell corrections to the free
energy, i.e., $F=F_{LD}$. The free energy with and without shell
corrections is plotted in panel (c), and  
the factor $({\cal I}_{LD}/{\cal I}_{SHL})^{3/2}$ is plotted in 
panel~(d).}
\label{f:fig6}
\end{figure}

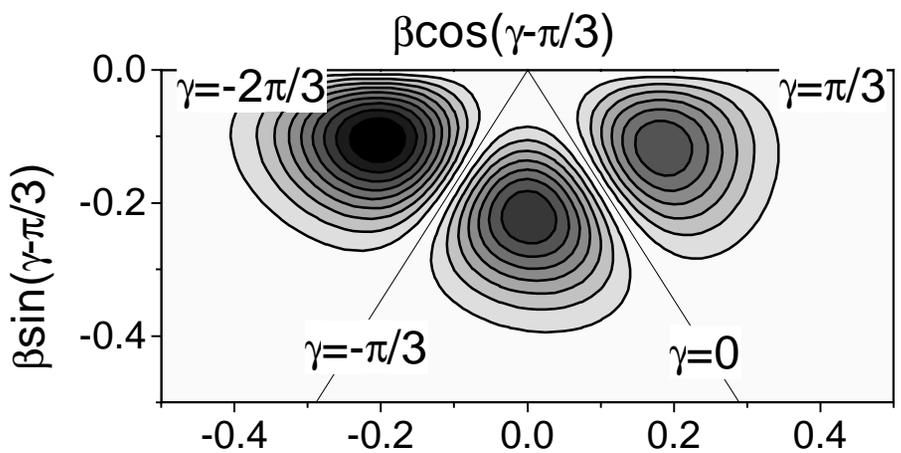
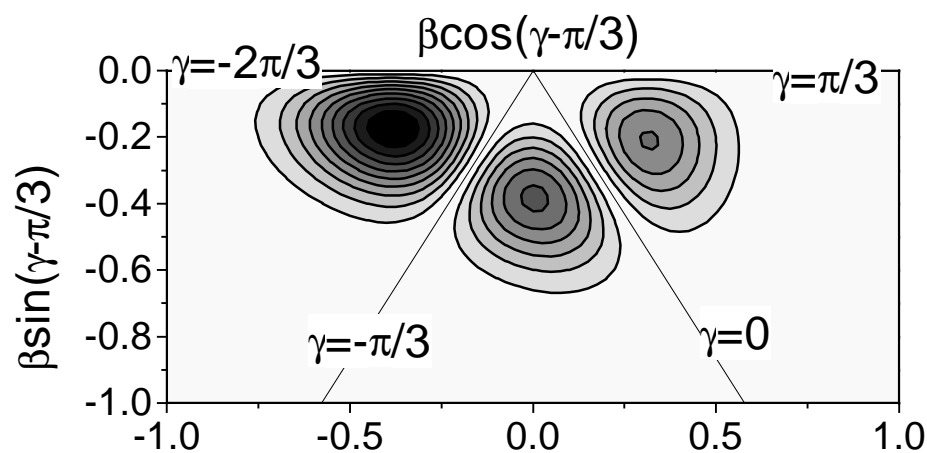
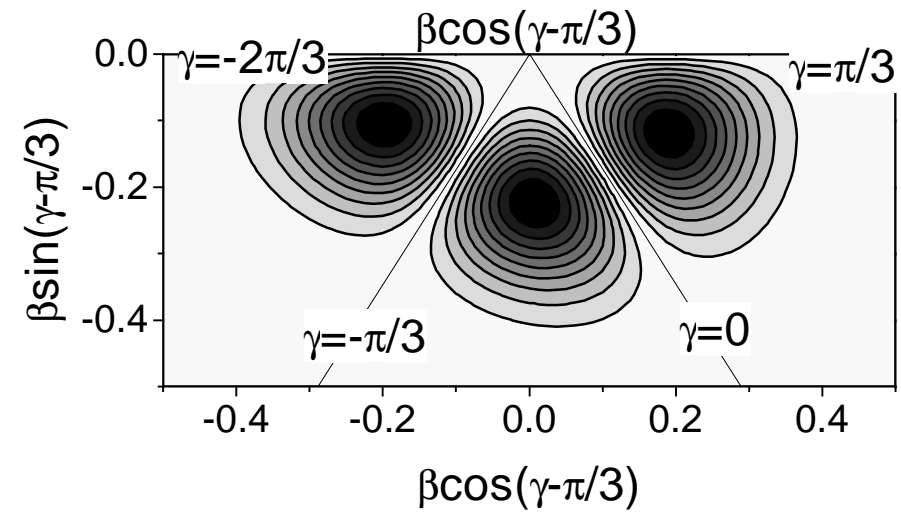
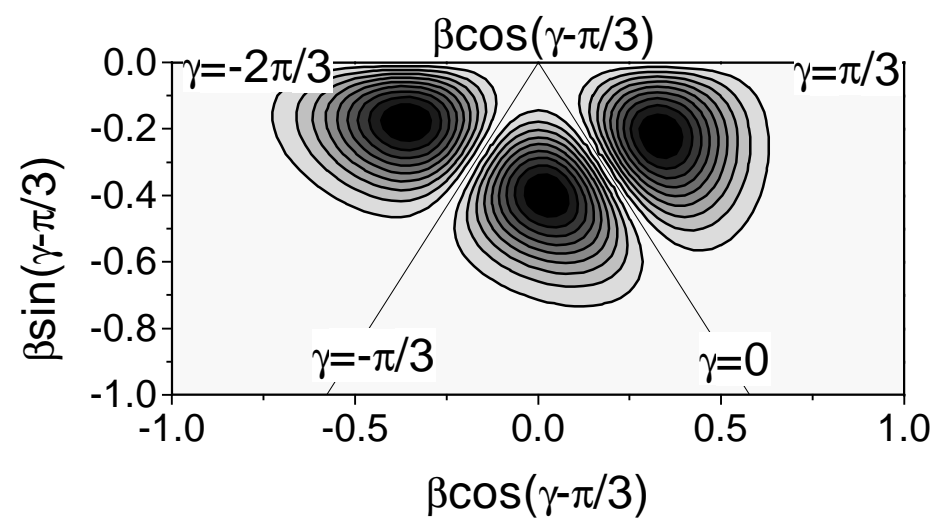
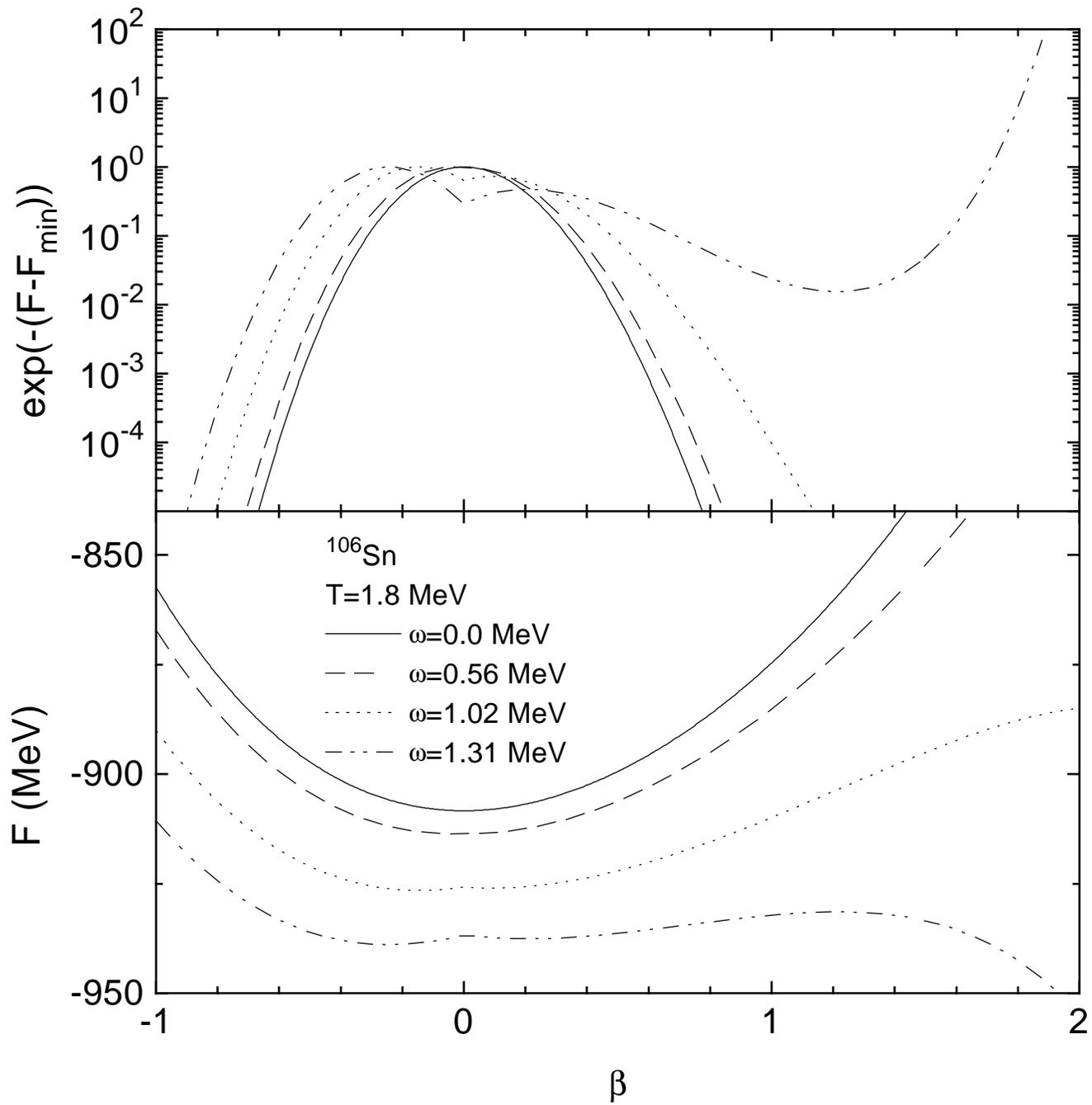
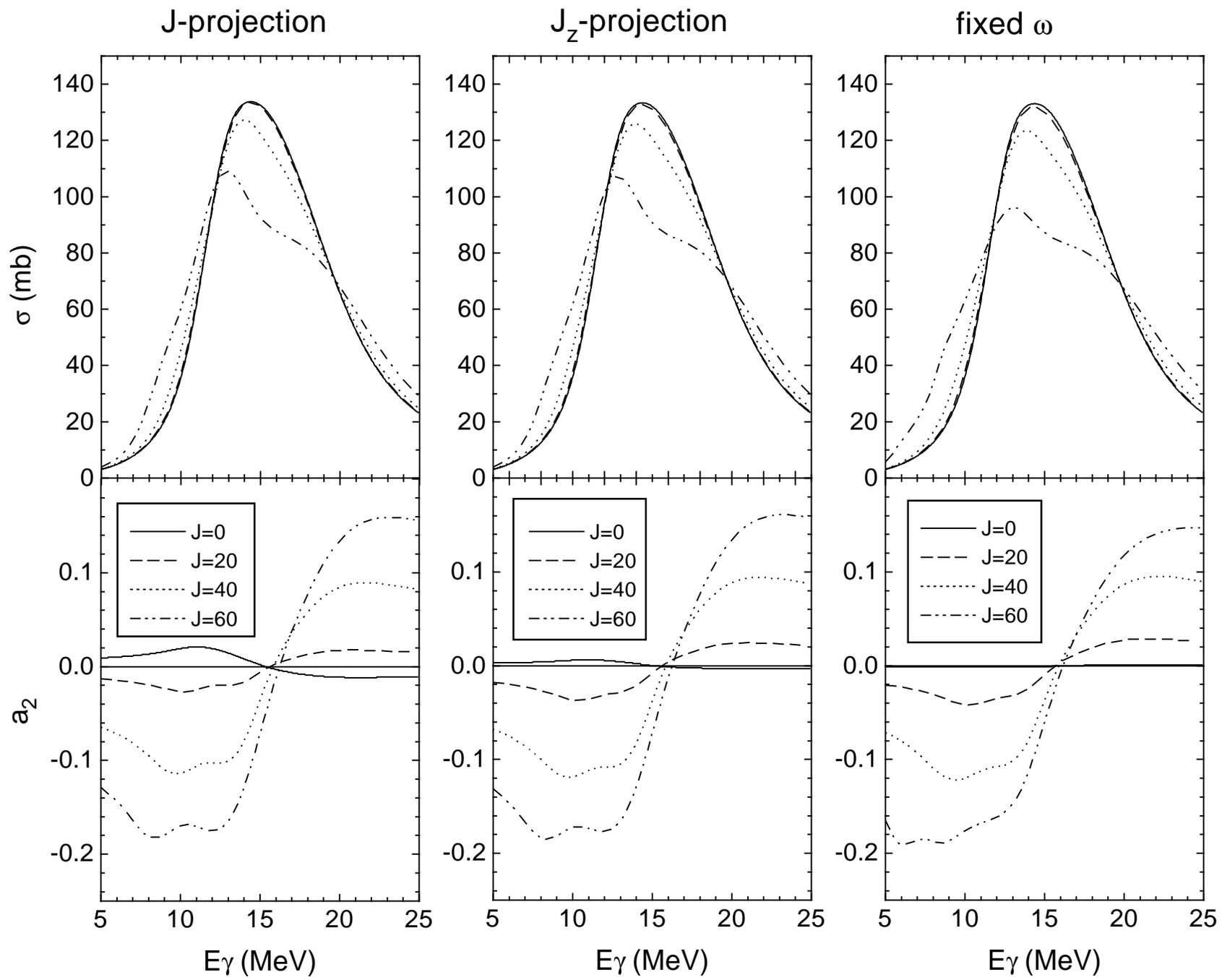
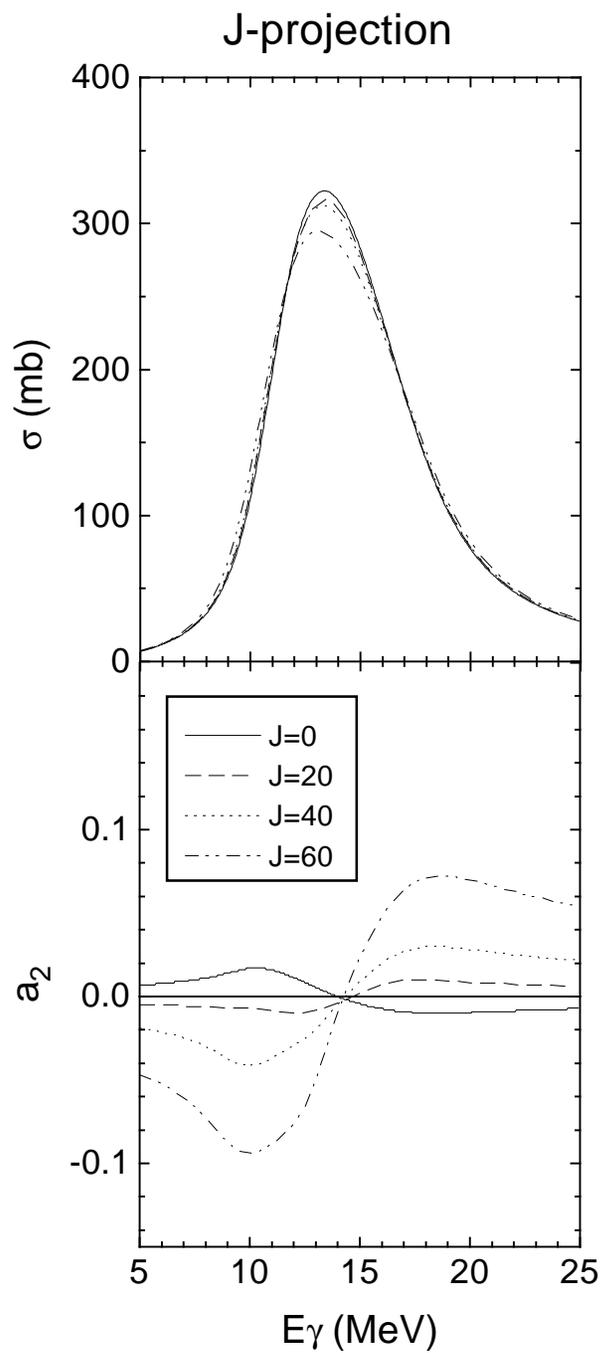
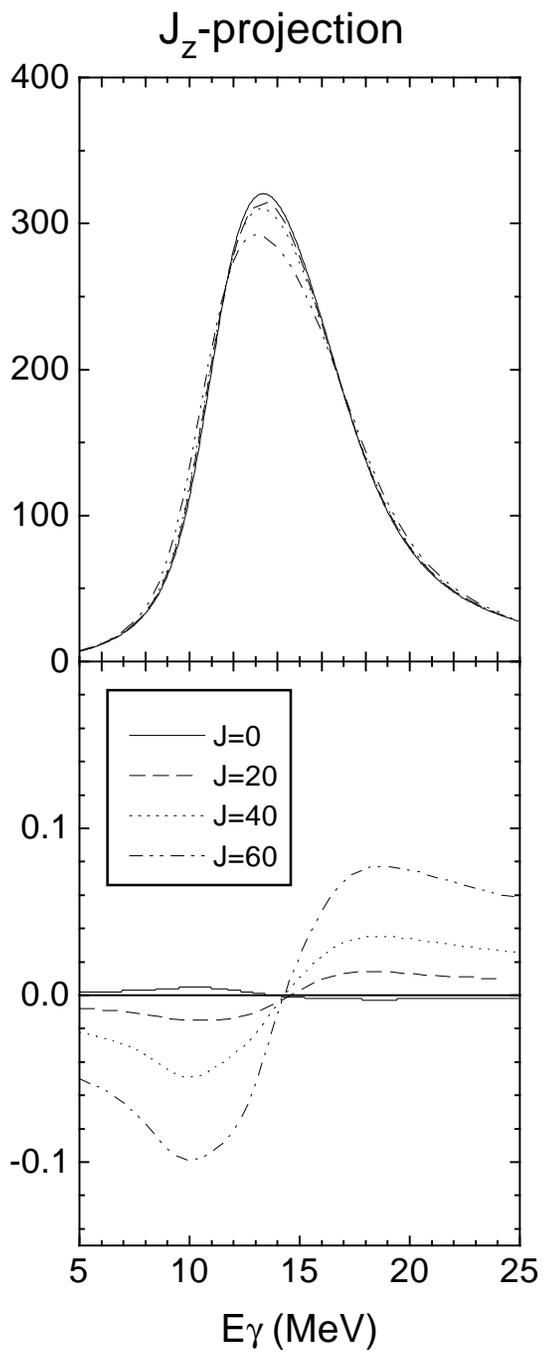
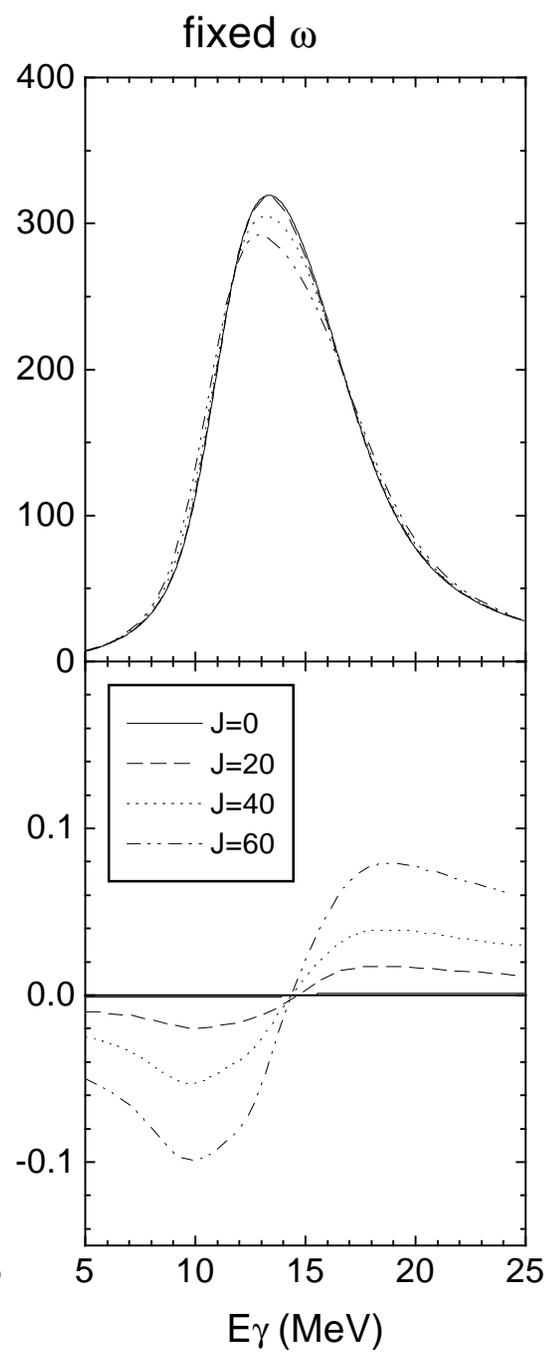
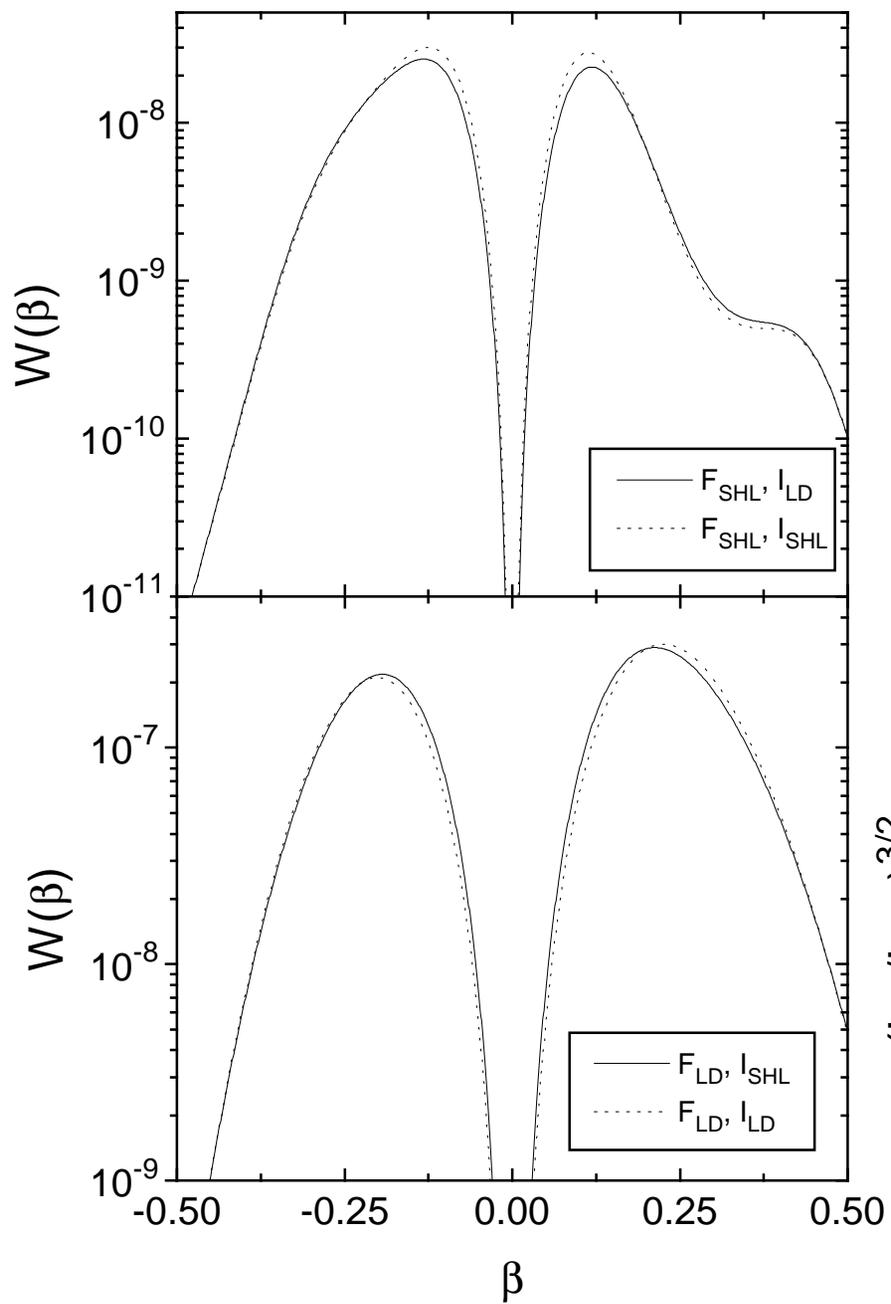
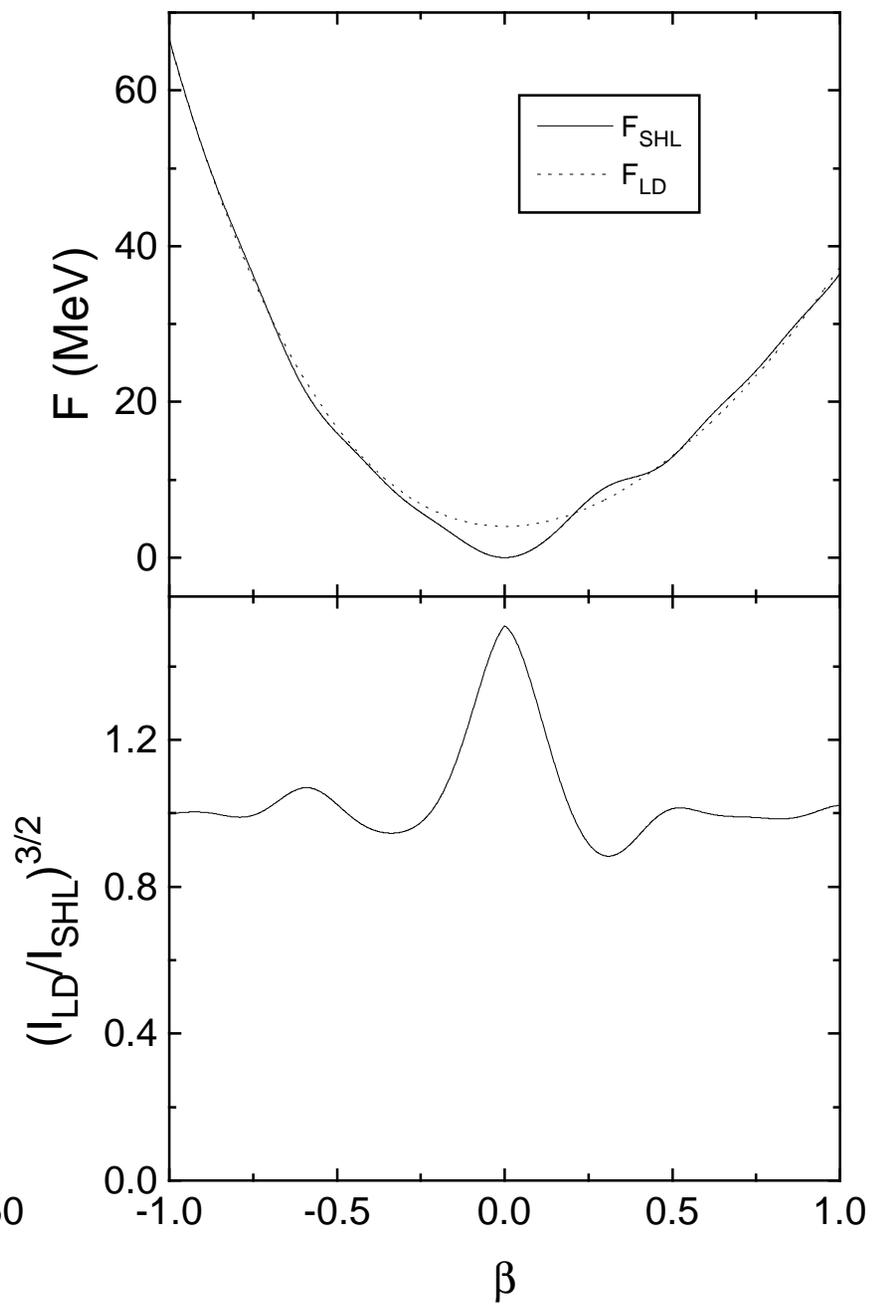
\end{document}